\documentclass[aps,preprint,showpacs,a4paper,superscriptaddress]{revtex4-1}

\usepackage{epsfig}
\usepackage{subfigure}

\newcommand{\bdm}{\begin{displaymath}}
\newcommand{\edm}{\end{displaymath}}
\newcommand{\benl}{\begin{equation}}
\newcommand{\be}[1]{\begin{equation}\label{#1}}
\newcommand{\ee}{\end{equation}}
\newcommand{\bea}{\begin{eqnarray}}
\newcommand{\eea}{\end{eqnarray}}
\newcommand{\ba}{\begin{array}}
\newcommand{\ea}{\end{array}}

\begin{document}

\title{Attractive interactions between like-oriented surface steps from an {\it ab-initio} perspective: role of the elastic and electrostatic contributions. }

\author{Giulia Righi}

\affiliation{Dipartimento di  Fisica, Informatica e Matematica (FIM), Universit\'a degli studi di
 Modena e Reggio Emilia and Centro S3 CNR-Istituto Nanoscienze, via Campi
 213/A, 41100 Modena, Italy}

\author{Anna Franchini}
\affiliation{Dipartimento di  Fisica, Informatica e Matematica (FIM), Universit\'a degli studi di
 Modena e Reggio Emilia}

\author{Rita Magri}
\affiliation{Dipartimento di  Fisica, Informatica e Matematica (FIM), Universit\'a degli studi di
 Modena e Reggio Emilia and Centro S3 CNR-Istituto Nanoscienze, via Campi
 213/A, 41100 Modena, Italy}
\email{rita.magri@unimore.it}

\begin{abstract}

In this work, we show how using the density functional approach, in which the electronic degrees of freedom are separated by the ionic ones, it is possible to individuate and separately study the elastic and electrostatic  step interactions, traditionally introduced in the literature as the only two kinds of step interaction expected at $T = 0$. We have applied the method to the technologically important GaAs(001) surface and found some unexpected results for the relatively short step distances accessible to the {\it ab-initio} approaches, contradicting those of the continuum models so far employed for the study of the elastic step interactions: (i) the sign of the step interaction depends on the step termination and is due to the electrostatic  interaction; (ii) the elastic interaction does not contribute to the step interaction, contrary to the common belief of a strong elastic repulsive interaction  between like-oriented steps. We show that this is due to the electron behavior. When considering only ion displacements and point-like steps as in the continuum theories, we recover the classical results and repulsive step elastic interactions; (iii) the experimentally observed $Ab$ step termination shows a weakly attractive step interaction whereby attractive step interactions between like-oriented steps on an unstrained surface are believed not to exist. The proposed method of separating elastic and electrostatic interactions for further analysis of their dependence on the configurational degrees of freedom can be extended to other defective situations.

\end{abstract}

\pacs{68.35.-p,68.35.Dv,68.35.Fx,68.47.Fg}

\date{\today}

\maketitle

%\newpage

%%%%%%%%%%%%%%%%%%%%%%%%%%%%%%%%%%%%%%%%
%\section{Introduction}
%%%%%%%%%%%%%%%%%%%%%%%%%%%%%%%%%%%%%%%%
\section{Introduction}
A problem that has attracted a great deal of interest and was never definitely solved is the existence of attractive interactions between steps on surfaces. The possibility to engineer the step density using only bottom-up concepts is very appealing due to the fundamental role steps play in surface faceting and in the formation and ordering of nanostructures\citep{ArcipreteACSNano2013} on the surface. Understanding how the step interaction depend on: (i) the kind of material and surface orientation, (ii) the surface reconstruction and step termination, and (iii) the surface growth conditions and post-growth processing is important. 
Controlled step edge density on ceria was used to distribute Pt catalysts since Pt atoms tend to decorate step edges\cite{Dvorak2016}. In reducible oxides the presence and density of steps influence the reducing oxide properties\cite{Miccio2016}. Furthermore, the formation of nanostructures such as dots and wires is driven by the underlying step configuration of the surface, thus  nanostructure arrangements are greatly influenced  by the position of surface steps. Step atomic  terminations play a role in the anchoring of nanoparticles to a surface.
Despite the importance of this issue, not much progress has been made in the past two decades. 

Most growth models and step dynamics simulations rely on parameters such as the step energy, the kink energy, and the step interaction energy. 
However, these parameters are at best simply estimated in their order of magnitude, and in most cases they are just adjusted to reproduce the desired behavior. 
The easiest way to model a sequence of steps on a surface is to consider a vicinal surface to a low-index facet. The vicinal surface orientation $\hat{n}$ has like-oriented steps that separate terraces of orientation $\hat{n}_0$.
The miscut angle $\alpha$ is the angle between $\hat{n}$  and  $\hat{n}_0$. The projected vicinal surface energy is commonly expressed as\citep{GruberJPhysChemSolids1967}
\begin{equation}\label{pfe}
\frac{\gamma(L)}{\cos (\alpha)} = \gamma^{\prime}(L)  = \gamma_0 + \frac{b(L)}{L}
\end{equation}
where $\gamma_0$ is the low-index surface energy, $b(L) = b_0 + g(L)$ is the step energy, and $L$ is the step distance related to the miscut angle by $L = \frac{h}{tan(\alpha)}$ with $h$ being the step height. The step energy $b(L)$ is given as the sum of the step self-energy $b_0$, i. e., the energy per unit length required to create a single isolated step, and $g(L)$, which is the step-step interaction that depends on the step distance $L$.
This expression relates the energy of the vicinal surface to the energy of the low-index surface without steps.

Eq.\ref{pfe}  is used in two ways  to derive the step parameters depending on the kind of theory adopted: 

(i) In the first approach, which is adopted in the framework of continuum theories,  the vicinal surface energy $\gamma$ is determined using Eq. \ref{pfe}  from a supposedly known $ \gamma_0$ after the expression for the step free energy $b(L)$ is explicitly determinated. Typically, a model of the step geometry  and of the force field generated by each step is assumed. The displacement field is then calculated and the step energy and step  interaction energy are derived. Using this approach, it was found that the main interaction energy is dipolar and  scales as $g(L)=\frac{K}{L^2}$ with the step distance $L$. At $T = 0$ the elastic interactions are assumed to be the strongest, and  they are repulsive ($K > 0$) for like-oriented steps, as they are on a vicinal surface. We term this theory the "dipole interaction model" (DIM).
The entropic effects at $T > 0$ add a further repulsive interaction with the same dipolarlike scaling with the step distance\citep{GruberJPhysChemSolids1967}.

(ii) The second way to use Eq. \ref{pfe} is adopted by the atomistic theories. These theories  make an atomistic model of the entire structure of the vicinal surfaces, and their free energy are calculated using inter-atomic potentials, or {\it ab-initio} calculations. The vicinal surface energy $\gamma (L)$ is directly calculated in the same way as the low-index $\gamma_0$. In this respect, the role of each single step cannot be isolated from the computed energy of the entire system. Consequently, Eq.\ref{pfe} is simply assumed to be correct, and the calculated energies are fit to it to extract  the step energy and, sometimes, the step interaction energy as well by assuming the DIM is valid.

 The idea of repulsive step interactions has since  been accepted and widely used to fit experimental data,  such as terrace width distributions from scanning tunneling microscopy (STM) images, in order to extract the step parameters. Increasingly refined theories have since been proposed, leading to the addition of logarithmic and multipolar corrections to the DIM\cite{MullerSurfSciRep2004}. However, none of these models has challenged in any decisive manner
 the dominant belief of a repulsive step interaction between like-oriented steps.

An initial challenge to the idea of repulsive step interactions came from the experiments. Attractive interactions between steps have been observed experimentally via STM measurements on Cu(100)\cite{Frohn1991}, on Ag(110)\cite{PaiSurfSci1994} and Si(113)\cite{SongPRB1995}, where they were revealed by a very inhomogeneous step density on the surface.  
Stimulated by these observations, modified models\cite{CiobanuPRB2003} and speculations\cite{RedfieldPRB1992} were advanced to explain the experiments, but a decisive reason for the existence of attractive step interactions between like-oriented steps was never found.

Up to now, continuum theories have addressed elastic interactions between widely spaced steps. Some authors have also advanced the possibility of electrostatic dipole interactions between steps as a possible source of attractivness \citep{LelargeEPL1997}, but clear experimental evidence or a theoretical investigation on their origin have never been proposed in the literature.  

Recently, we fit a number of {\it ab-initio} calculated surface energies of vicinals to a $\beta_2 (2 \times 4)$ reconstructed   GaAs(001) surface for different step edge configurations and miscut angles to Eq.\ref{pfe}  using the DIM\cite{Magri2014, Erratum}.  $\beta_2 (2 \times 4)$ reconstruction is observed over a wide range of conditions employed in the growth of compound semiconductor devices.
 The fit revealed, surprisingly, that on this surface the interaction between steps may be attractive, at least for the short step distances accessible by {\it ab-initio} calculations. Thus, we are  in the unique position to explore what may determine the attractive interaction between steps, and the issue of electrostatic step interactions, using a computational tool in which the valence electrons are described quantum-mechanically. This is an improvement over both the classical continuum theories and the semiempirical inter-atomic potentials used so far.
 
\section{Method}  
The surfaces, both $\beta_2$ and its vicinals, are modeled through periodic slabs. 
The vicinal surfaces are described through a regular sequence of equally distanced straight step lines oriented along the $[\overline{1} \  1 \  0]$ surface direction
separated by  $\beta_2 (2 \times 4)$ reconstructed terraces of variable dimensions.
The steps, termed $A$ steps, have been found to have the lowest step energy\citep{HellerPRL1993}. Their direction is parallel to the direction of the As dimers of the  $\beta_2 (2 \times 4)$ reconstruction. 

We consider here two different step atomic configurations:
one, termed $Aa$, is slightly Ga-richer and the other, termed $Ab$, is slightly As-richer compared to the  $\beta_2 (2 \times 4)$ surface. 
The $Ab$ step configuration is the lowest-energy one\citep{Erratum} and was observed by Kanisawa et al.\citep{KanisawaPRB1996} on GaAs(001) (2x4) surfaces using high-resolution STM.

\begin{figure}
\centering
\includegraphics*[scale = 0.5]{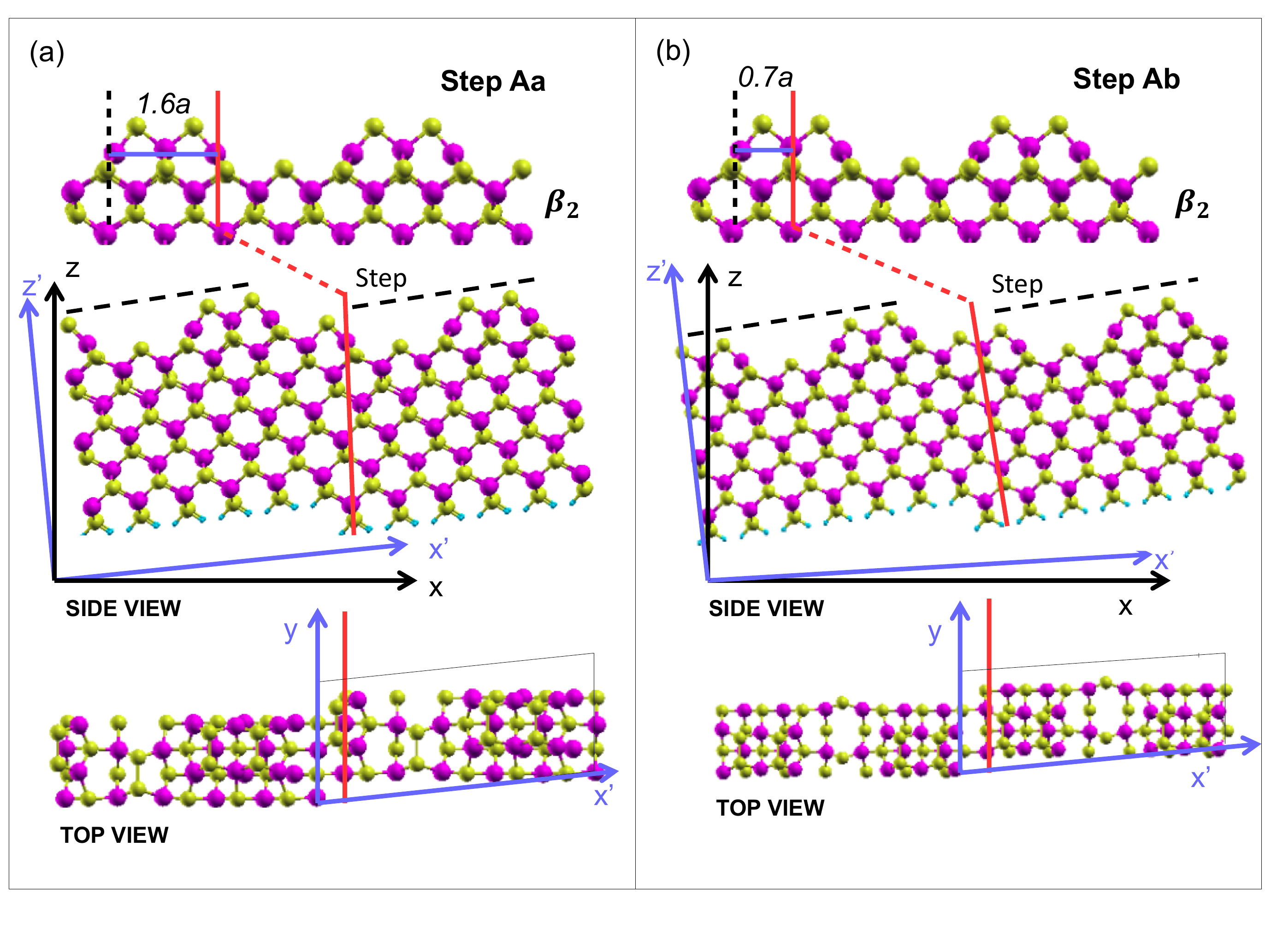}
\caption{(Color online) Construction of the $Aa$ and $Ab$ vicinals from the $\beta_2$ surface. Starting from a given reference atom (dashed black line), a vertical plane (red line) is drawn. The periodicity along the $x$ direction is changed and the structure is rotated. Side and top views of the vicinal surfaces: (a) Step $Aa$ and (b) Step $Ab$. The thin solid lines in the top views indicate the surface unit cells. Purple, Ga; yellow, As; blue small balls, pseudo-hydrogens.  }
\label{Fig1}
\end{figure}

The vicinal surfaces are constructed using optimized $\beta_2 (2 \times 4)$ surface building blocks.
The atoms belonging to the $\beta_2 (2 \times 4)$ surface slab are labeled. The vicinal surfaces with steps $Aa$ and $Ab$ are then obtained from the $\beta_2 $ surface by cutting it with a (1\ 1\ 0) plane passing through a given atomic plane, as shown in Fig. \ref{Fig1}. The vicinal surface unit cell is then constructed by extending the lattice vector along the [1\ 1\ 0] direction. 
The step structural motif develops naturally by the application of the system periodicity  along the [1\ 1\ 0] direction. 
The dimensions of the considered structures are $y = 2a$ along  [$\overline{1}$\ 1\ 0] (as in $\beta_2$) and they range
from $5.5 a$ (miscut angle $\alpha = 7.33^{\circ}$) to $17.5 a$ ($\alpha = 2.31^{\circ}$) and from $6.5 a$ ($\alpha = 6.21^{\circ}$) to $18.5 a$ ($\alpha = 2.19^{\circ}$) along the $x^{\prime}$ direction for the vicinal surfaces with $Aa$ and $Ab$ steps, respectively.  $a$ is the surface lattice parameter, $a=\frac{a_{0}}{\sqrt{2}}$ with $a_{0}$ the calculated GaAs bulk lattice parameter. The atoms forming the vicinal surface are in correspondence with those of the $\beta_2 $ surface (albeit with repetitions since the vicinal unit cell is longer than that of $\beta_2$). 
The resulting structures are shown in Fig.\ref{Fig1} for the case of the largest $\alpha$ angle.

We calculate the formation energies per unit area of the vicinal surfaces as:

\begin{eqnarray}\label{ef}
\gamma_{i} (\alpha,\mu_{As}) &=& \frac{\left( E_{i,\alpha} - n_{\mathrm{Ga}} \mu^{bulk}_{\mathrm{GaAs}} 
+ ( n_{\mathrm{Ga}} - n_{\mathrm{As}}) \mu^{bulk}_{\mathrm{As}} \right)} {S} \nonumber \\
&& + \frac{( n_{\mathrm{Ga}} -n_{\mathrm{As}}) \Delta \mu_{As} }{S}.
\label{ef}
\end{eqnarray}
In Eq. \ref{ef},
$E_{i,\alpha}$ is the total energy of the vicinal surface slab where  $i = Aa\  \mathrm{or}\  Ab$, and $\alpha$ is the miscut angle. $S$ is the surface unit cell area, and $n_{\mathrm{Ga}}$ and $n_{\mathrm{As}}$ are the number of Ga and As atoms in the system, so that  $n_{\mathrm{Ga}}-n_{\mathrm{As}}$ defines the stoichiometry. 
$\mu_{GaAs}^{bulk}$ is the energy required to form a Ga and As pair in bulk GaAs.  The dependence of the surface energy on the growth conditions is represented by  $\Delta\mu_{As}$, which is the change of the As chemical potential $\mu_{As}$  relative to its value in the bulk rhombohedral As metal. Thus, in Eq. (\ref{ef}) the energy of the vicinal surface depends only on $\Delta\mu_{As}$ since for the surface in equilibrium with its bulk, $\mu_{Ga}+\mu_{As} = \mu_{GaAs}^{bulk}$. 
In what follows, we will use for simplicity $\Delta\mu_{As}=0$, corresponding to extremely high As coverages, which are not appropriate for the $\beta_2$ surface. However, since we are interested only in the trends of the vicinal surface energies with the step distance, our conclusions are unaffected by the value of the As chemical potential, which causes only an $\alpha$-independent rigid shift of all the surface energies. 

The step distances we consider are found on the mounds formed by unstable growth on GaAs (001) surfaces ($\approx 7$ nm). They were also observed by Pashley et al.\cite{PashleyAPL1991} on a $2 \times 4$ vicinal (001) GaAs surface cut 2$^{\circ}$ toward (111)A via STM.  Perfectly straight steps over distances of the order of 100$\AA$ with terrace widths variable from 3 to 8 ($2 \times 4$) unit cells were seen. Thus, the vicinals we consider here are representative of actual experimental situations. 

We use the density functional theory (DFT) as implemented in the Quantum Espresso \cite{QE-2009,QE-2017} suite. The calculation by DFT of surface step energies is very complicated since surface energies are tiny quantities given by the difference between  two usually very large total energies. In addition, the step energetics are derived from a comparison of surface energies calculated for similar surface terminations. In our reciprocal space based {\it ab-initio} approach, we model the surface through material slabs in the vacuum. It was shown that in the case of titania surfaces, the surface energies have an oscillating behavior with the number of atomic layers making up the slab\cite{Hardcastle2013}. Thus, the precise values of the surface step energies and step interaction energies depend on the surface slab model and on the calculation parameters.  
In this work, however, we are concerned with the sign of the step interaction, not its precise values. The sign of the step interaction depends on the trend of the surface energy with the step distance. Since the total energies we compare are calculated using the same slabs with the same number of layers, with the backside saturated by the same kind of atoms (whose energy is then subtracted, taking account of the back steps) and the same calculation parameters, most of the systematic errors get cancelled. Moreover, we will consider differences of surface energies (i.e, the energy of the vicinal surfaces minus the energy of the corresponding low index surface) which should further cancel any residual systematic error.  Indeed, we find that while the precise values change if we use  different pseudopotentials or a different number of layers in the slabs, the trend does not change significantly.
Further technical details and the convergence tests are reported in the supplemental material. 

\section{Calculated step interaction}
\begin{figure}
\centering
\includegraphics*[scale = 0.6]{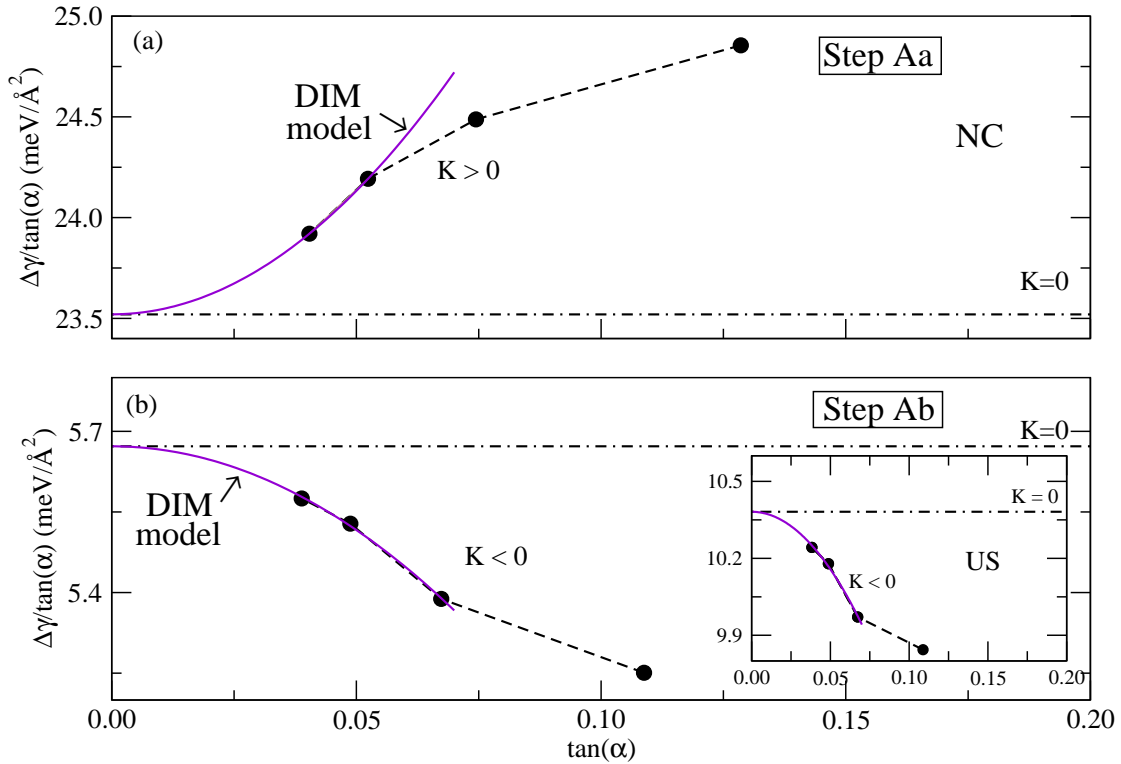}
\caption{(Color online)  Calculated $f(\tan \alpha)$ for (a) step $Aa$ and (b) for step $Ab$. The calculations are performed using norm-conserving (NC) pseudopotentials. In the inset we show the results obtained for step $Ab$ using instead ultrasoft (US) pseudopotentials. }
\label{Fig2}
\end{figure}

In  Fig. \ref{Fig2} we plot $\Delta \gamma /\tan (\alpha)$, where $\Delta \gamma = (\gamma^{\prime} - \gamma_0)$ with $\gamma^{\prime}$ the calculated projected surface energies and $\gamma_0 = \gamma_{\beta_2}$. In this procedure, the smaller the values of $\alpha$ are, the larger is the scale. If there were random relevant errors in the calculations, we would obtain random points, but instead the obtained values have a reasonable behavior. 
To extract the step parameters, we use approach (ii) of Sec. I, and refer to Eq. \ref{pfe}. The DIM would read:

\begin{equation}\label{pfe_mod}
f(\tan \alpha) = (\gamma^{\prime} - \gamma_0)/\tan (\alpha) = b_0/h + (K/h^{3})\tan^2 (\alpha)
\end{equation}

which has a parabolic trend with a constant $K$. The absence of step interaction ($K = 0$) is represented by a horizontal straight line. An increasing $f$ would mean $K>0$, i. e., a repulsive step interaction, while a decreasing $f$ is related to $K<0$ and an attractive step interaction. Thus, we see clearly that our calculations produce a repulsive step interaction for step $Aa$  and a weakly attractive one for step $Ab$. 
Apart from the attractive step interaction, we can also see that our calculated values do not follow the simple DIM expression with a constant $K$ (purple lines in Fig 2). This behaviour could be due to the too short step distances in the cases of  terraces formed by only one or two  $\beta_2$ unit cells and the overlap of the step effects.

Our results show that attractive step interactions may exist. The question arises about how this result relates to the continuum theories, which always predict a repulsive elastic interaction between like-oriented steps.

\section{Comparison between continuum models and the {\it ab initio} approach}
First, we observe that the continuum theories refer mainly to elastic interactions. Some authors have advanced the idea of  electrostatic dipole step interactions\cite{LelargeEPL1997,RedfieldPRB1992,Magri2014}, but their origin and their relation to the elastic step interactions have never been discussed. 
Here we show that this can only be done within a theory capable of separating the electronic degrees of freedom from the ionic ones. 

\begin{figure}
\centering
\includegraphics*[scale = 0.5]{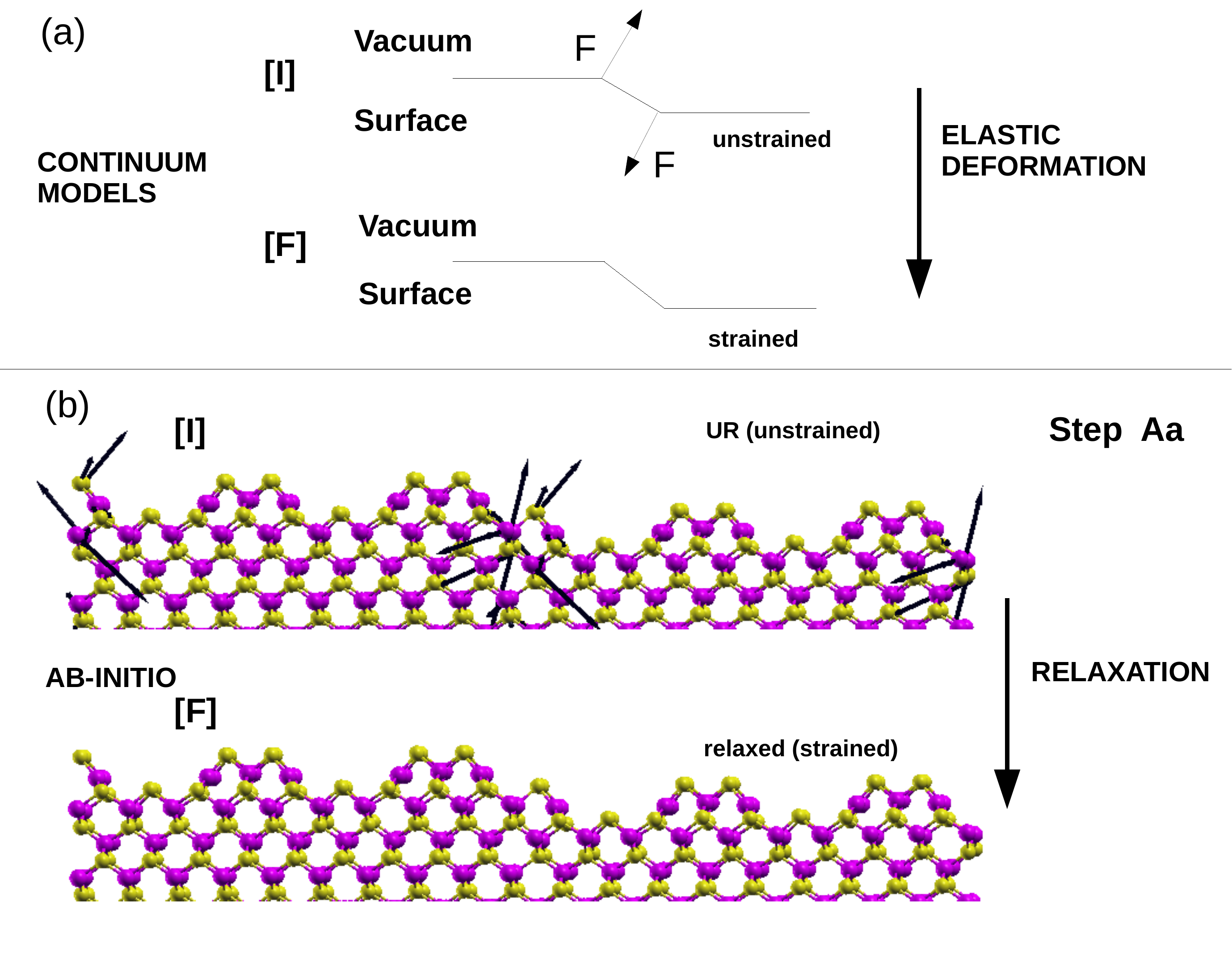}
\caption{(Color online) (a) Classical formalism: [I], step dipolar forces on the unstressed surface; [F], final strained surface configuration. (b) The same situations for the DFT formalism. Purple, Ga; yellow, As. }
\label{Fig3}
\end{figure}

We start first with a comparison between the continuum models and the DFT approach.  The continuum models begin with a model of step forces, such as that shown in Fig. \ref{Fig3}a. Starting from this initial situation,
the step energy and the step interaction energy are calculated from the work performed by the step forces and the corresponding displacement fields. The coupling between steps so derived is therefore through the elastic field.

Likewise we consider the vicinal surfaces obtained by pasting together the different pieces of  $\beta_2$. We initially keep each  ion  in the same position as in the $\beta_2$ surface. Only across the junctions do atoms of the original $\beta_2$ surface find themselves near differently labeled atoms (the kind remains the same). The atom coordination remains as it is in the $\beta_2$ slab. Only on the surface atomic layer are new structural motifs at the steps developed. 
These differences define the structural "defect" acting on the otherwise optimized $\beta_2$ surface. This "defect"
 causes a charge redistribution and the formation of forces very well localized at the step region.
The calculated forces for the vicinal surface with $Aa$  steps  are shown in  Fig. \ref{Fig3}b, configuration [I]. We notice the similarity to the initial situation [I] of Fig. \ref{Fig3}a.\\
In both cases, the step forces  drive the material displacement. In the {\it ab-initio} approach, the process is commonly termed structural relaxation and leads to an energy lowering.  In the continuum theories, the surface energy of the elastically displaced material [configuration [F]  of Fig. \ref{Fig3}a] is considered higher than the surface energy of configuration [I], which is taken to be the energy of the flat surface at its equilibrium.  
 However, we notice that  the configurations [I] of Fig. \ref{Fig3}(a) are not at their equilibrium because of the presence of forces acting upon them, so an additional piece of energy has to go in creating situation [I] from the true equilibrium flat surfaces.

To illustrate the point, we decompose $\Delta \gamma(\alpha)=\gamma^{\prime}(\alpha)- \gamma_{\beta_2}$ as:

\begin{equation}\label{deltagamma}
\Delta \gamma(\alpha) = \gamma^{\prime}(\alpha)-\gamma^{UR}(\alpha) +\gamma^{UR}(\alpha)- \gamma_{\beta_2}. 
\end{equation}

 $\gamma^{UR}(\alpha)$ (where $UR$ denotes unrelaxed) is the projected surface energy of the vicinal surface in configuration [I] of Fig. \ref{Fig3}b, thus
 the quantity $\Delta \gamma^{UR}(\alpha) = \gamma^{UR}(\alpha) - \gamma_{\beta_2}$ is the energy  per unit area required to  construct the nonequilibrium vicinal surface in configuration [I] from the $\beta_2$ surface, i.e., the positive energy due to the insertion of the "defect".
Since the ions have not yet been moved from the positions they have in the $\beta_2$ surface, this piece of energy represents only the electronic response (electronic charge redistribution) induced by the presence of the steps that originate the step forces depicted in Fig. \ref{Fig3}b [I].

 The term $\gamma^{\prime}(\alpha)-\gamma^{UR}(\alpha) = \gamma_{rel} $ is the negative relaxation energy.
 This energy is related to the ion displacements and the corresponding electronic charge rearrangements. We identify $ \gamma_{rel}$  with $-\gamma^{EL}$,  the positive elastic ($EL$)
material deformation of the continuum models. This identification extends the definition of strain energy as intended so far within the {\it ab initio} approaches, where it has always been defined and calculated in relation to  lower energy, high symmetry reference systems, as in the case of epitaxial strains \cite{Zhang2015, King-Smith1994}, or when calculating force constants \cite{Kresse1995}. In all these cases, the electronic charge rearrangement is considered together with the ion displacements, and are integral part of the force constants.
Here, however, we refer $\gamma_{rel}$ properly to the higher energy reference system, since the existence of forces (necessary to cause the displacements) implies a starting nonequilibrum  situation.
This identification allows us to extend also the notion of {\it ab initio} elasticity to inhomogeneous strain fields.

The continuum models, while calculating the elastic energies using step forces, refer instead the result to the lower-energy equilibrium system $\gamma_0$. In doing so, they account for all the difference $\gamma^\prime - \gamma_0$. In this scheme, there is no space for any electrostatic step interaction, which is considered to be an unrelated additional interaction for which no method has yet been devised for its estimate.

\section{Extracting elastic and electrostatic step interactions}

By identifying the relaxation energy of the system from the higher-energy unrelaxed "step defect" configuration with the opposite of the elastic strain energy, we have in fact changed the reference system and neglected the energy difference between the optimized $\beta_2$ surface and the UR vicinal at a higher energy, as is done in the continuum models.  It is easy in this context to extract also the electrostatic ($ES$) contribution under the assumption that only these two kinds of step interation (elastic and electrostatic) play a role at $T = 0$. Writing $b_0 = b_0^{EL}+b_0^{ES}$ and $g(L)= g^{EL}(L) + g^{ES}(L) $ for $\gamma^{\prime}-\gamma_0$ in Eq. \ref{pfe}, we write $\gamma^{\prime}-\gamma_0 = \gamma^{EL}+\gamma^{ES}$,
  where both terms may contribute to the step energy and the step interaction.
  Comparing this expression with Eq. \ref{deltagamma} using $ \gamma_{rel} = -\gamma^{EL}$ we get $\gamma^{ES} = \Delta \gamma^{UR}+2\gamma_{rel}$, an expression for the electrostatic energy.
Now we are in a position to evaluate and compare separately both terms and see how they depend on the step atomic configuration.

\section{Discussion of results}

\begin{figure}
\centering
\includegraphics*[scale = 0.5]{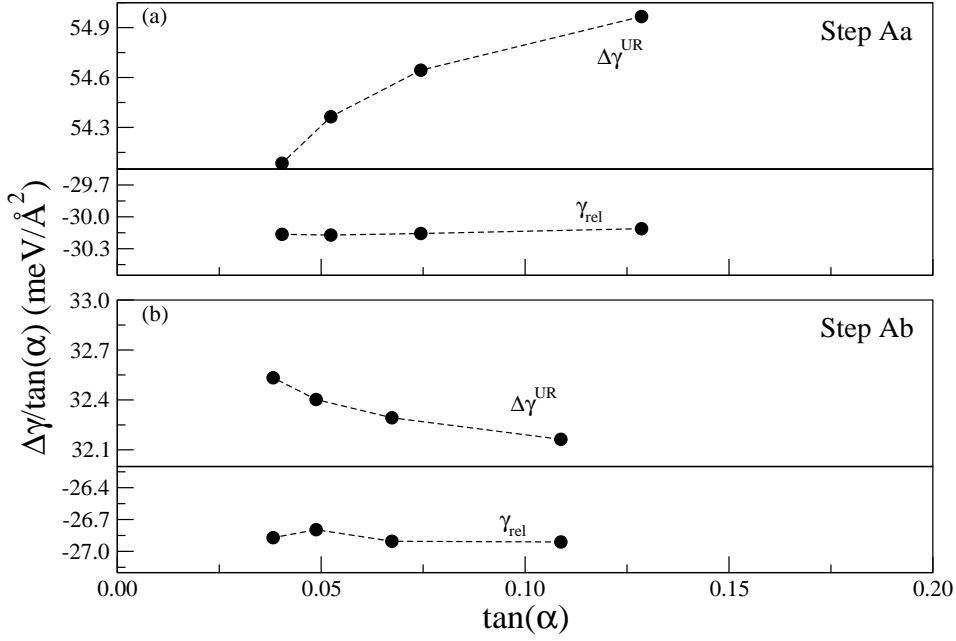}
\caption{Trends of $\Delta \gamma^{UR}/\tan( \alpha)$ and $\gamma_{rel}/\tan( \alpha)$ as a function of $\tan (\alpha)$ for (a) stepAa, and (b) stepAb. The dots are the calculated contributions. }
\label{Fig4}
\end{figure}

\begin{figure}
\centering
\includegraphics*[scale = 0.5]{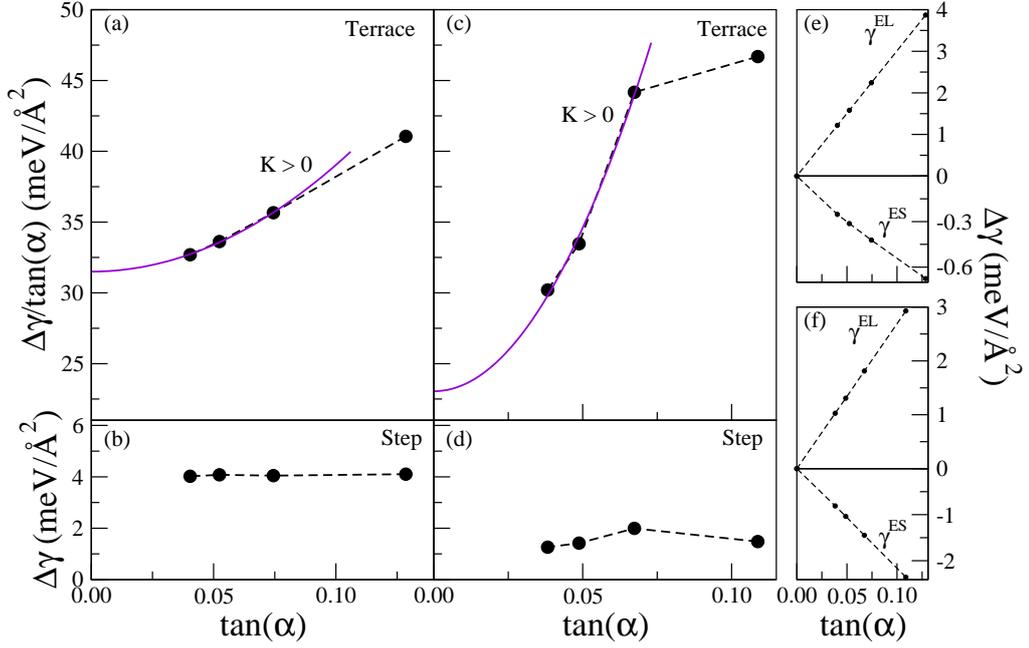}
\caption{(Ionic contribution  $\gamma^{EL}_{ion}$ to the elastic energy divided by tan($\alpha$) of  the terrace ions for the vicinal surface having (a) steps $Aa$,  and (c) steps $Ab$; contributions to the ionic elastic surface energy due to the step regions for steps $Aa$ (b) and step $Ab$ (d); the elastic $\gamma^{EL}$ and electrostatic $\gamma^{ES}$ contribution to the difference $\Delta \gamma = \gamma^{\prime} - \gamma_0$ for the vicinals with (e) steps $Aa$ and (f) steps $Ab$. }
\label{Fig5}
\end{figure}

In Fig. \ref{Fig4} we plot and compare $\Delta \gamma^{UR} /\tan( \alpha)$ and $\gamma_{rel}/ \tan (\alpha)$ in (a) and (b) for the steps $Aa$ and $Ab$, respectively. Since  $\gamma_{rel}/\tan(\alpha)$ is independent on $\alpha$ (then $K$=0), we find that the elastic interactions do not contribute to the step interaction, contrary to what is commonly believed. To better understand this result,  we have estimated the contribution to the  elastic energy due to the classical ion displacements (ions are classical objects in the DFT codes)
 by calculating the work done by the ab-initio forces during the ionic relaxation.  Using the BFGS alghoritm \cite{fletcher1987, billeter2003} implemented in the PWSCF code,  we have calculated:
 
 \begin{equation}
\gamma^{EL}_{ion} = \sum_i \sum_j F_{ij}\cdot (s_j(i)-s_j(i-1)) 
 \end{equation}  
where $i$ runs over all the iterations from configuration $UR$ to the final optimized step geometry, and $j$ runs over all the ions. $s_j$ are the displacements of ion $j$. The forces $F_{ij}$ are  the total forces, and they are the sum of the contributions due to both the ion-ion interaction and the electron-ion interaction.
The results for both kinds of steps are given in Figs. \ref{Fig5} (a) and 5(c) for step $Aa$ and $Ab$, respectively, where we report only the contributions to $\gamma^{EL}_{ion} / \tan \alpha$ due to the terraces (complete $\beta_2$ units), that is, the quantity 
$\gamma^{EL}_{ion} \left( \mathrm{terrace} \right) /S^T$, where $S^T$ is the portion of the area under the terrace.
The contributions to $\gamma^{EL}_{ion}$ due instead to the step regions (under the new surface structural motifs), wide $1.5 a$ and $2.5 a$ for steps $Aa$ and $Ab$, respectively, are given in (b) and (d). These contributions are calculated as $\gamma^{EL}_{ion} \left( step \right) /S^S$, where $S^S$ is the portion of the surface area under the step region. Obviously, $S^T + S^S = S$ is the total surface. 

The terrace contributions can be compared with the continuum theories assuming steps are lines of point forces without a physical extension as in the papeer by Marchenko and Parshin\citep{MarchenkoSovPhysJETP1980}. As in that work, we find a repulsive trend, i. e. if  (i) we only consider the classical objects (ions), (ii) we neglect the true spatial extension of the  steps, and (iii) we refer to the flat $\beta_{2}$ surface, then we recover the classical result. Thus, this shows that the repulsive elastic behavior between like-oriented steps is based on the set of assumptions made by the classical continuum theory.

As for the step region contribution to  $\gamma^{EL}_{ion}$, we see that it is insensitive to the step distance, as it should be.  This constant step contribution is part of the step energy, the energy of the isolated step, and thus it does not contribute to the step interaction energy.

Thus, the fact that the strain does not contribute to the step interaction as shown for $\gamma_{rel}=-\gamma^{EL}$ in Fig. \ref{Fig4} is ultimately due to the electron response to the ion displacements,  which  
 counteracts the energetic effects of the ionic motion for what concerns the step interaction.  

Finally, we compare the relative strenghts of the elastic and electrostatic contributions to  $\gamma^{\prime}-\gamma_0$ in Fig. \ref{Fig5}(e), and \ref{Fig5}(f), where we 
 show $\gamma^{EL}$ and $\gamma^{ES}$ as a function of $\tan (\alpha$).  The electrostatic interaction is much stronger for  step $Ab$, where it is of the same order of the elastic one, than for the step $Aa$. Moreover, we can see that
the sign of the step interaction is related to the concavity of the electrostatic contribution to $\gamma^{\prime}-\gamma_0$, i. e., repulsive for step $Aa$ and slightly attractive for step $Ab$. Indeed, the fit of the values reported in Fig. \ref{Fig5}(e), and \ref{Fig5}(f) using Eq. \ref{pfe} and the DIM approximation produce the values reported in Table I.

\begin{table}
\caption{ Values obtained through the fit of the elastic $EL$ and electrostatic $ES$ contributions to the surface energy difference:  the step energy $b_0$ (in meV/\AA) and the step interaction energy constant $K$ (in meV$\cdot$\AA) for vicinals having steps $Aa$ and steps $Ab$.}
\begin{tabular}{ccccc}
\hline
\hline
& \multicolumn{2}{c}{StepAa} &   \multicolumn{2}{c}{StepAb}\\
\hline
& $b_{0}$ & $K$ & $ b_{0} $& $K$\\
\hline
& & & &\\
$\gamma^{EL}$ & 84.64 & -88.84 & 75.29 & 132.72\\
& & & &\\
$\gamma^{ES}$ & -17.03 & 1104.64 & -59.63 & -765.54\\
& & & &\\
\hline
\hline
\end{tabular}
\end{table}

The values reported in the table show that the $K$ values of the elastic step interaction $\gamma^{EL}$ are negligible when compared to those corresponding of the electrostatic step interaction $\gamma^{ES}$. Thus, we can see that the sign of the step interaction is mainly decided by the electrostatic contribution.

\section{Conclusions}

In conclusion we have carried out an extensive and critical examination of the nature of the elastic and electrostatic interactions between like-oriented steps on a vicinal surface to  GaAs (001) using an {\it ab initio} approach based on the density functional theory. We have compared the {\it ab initio} approach to the usually employed classical approaches based on continuum elasticity. Continuum elasticity approaches describe only the elastic deformation of the surface and its associated energy. Their derived elastic step interaction energy is always repulsive for like-oriented steps on an unstressed surface.
We show that these approaches neglect  the "excitation" energy due to the step presence relative to the unstepped surface, which is the cause of the material deformation itself, and have identified the {\it ab initio} strain relaxation energy  with the elastic energy of the classical continuum theories (with an opposite sign). The change in sign is equivalent to a change in the reference system from the "excited" system (a surface with unrelaxed steps) to the flat surface at its equilibrium.

The calculations of the surface energies, their elastic and electrostatic contributions, and the step interaction energies have shown many unexpected trends. The results change most of the current knowledge about the nature of the step interaction at $T = 0$.
We have  found that (i) the step interaction energy and sign depend strongly on the step termination and are due to the electrostatic  interaction between steps; (ii) the elastic interaction does not contribute to the step interaction, contrary to the common belief of a strong elastic repulsive interaction  between like-oriented steps. We have shown that this is mainly due to the electron response to the ion displacements. When we consider only the ions displacements and neglect the spatial extension of the steps, assimilated to  point lines as in the continuum theories we recover the classical results and repulsive step interactions; (iii) the experimentally observed step termination, step $Ab$, shows a weakly attractive step interaction whereby attractive step interactions between like-oriented steps on an unstrained surface are believed not to exist.

These results, apart from their fundamental value, show that the strength and sign of the step interaction depend strongly on the step termination, and they allow for a better analysis of the reasons underlying the interaction itself, by linking the structural step motifs to distinct  and different elastic and electrostatic contributions (albeit lots of work has still to be done on this point). This work constitutes a step forward,   
paving the way to possibly engineering step interactions via surface atomic manipulations.

\begin{acknowledgments}
The authors would like to thank the Supercomputing Center CINECA, Bologna, Italy, for providing computing time under the two projects IscrC-ELETTO and IscrB-UNDEFEAT.
\end{acknowledgments}

\bibliography{Steps}

\providecommand{\noopsort}[1]{}\providecommand{\singleletter}[1]{#1}%
\begin{thebibliography}{25}%
\makeatletter
\providecommand \@ifxundefined [1]{%
 \@ifx{#1\undefined}
}%
\providecommand \@ifnum [1]{%
 \ifnum #1\expandafter \@firstoftwo
 \else \expandafter \@secondoftwo
 \fi
}%
\providecommand \@ifx [1]{%
 \ifx #1\expandafter \@firstoftwo
 \else \expandafter \@secondoftwo
 \fi
}%
\providecommand \natexlab [1]{#1}%
\providecommand \enquote  [1]{``#1''}%
\providecommand \bibnamefont  [1]{#1}%
\providecommand \bibfnamefont [1]{#1}%
\providecommand \citenamefont [1]{#1}%
\providecommand \href@noop [0]{\@secondoftwo}%
\providecommand \href [0]{\begingroup \@sanitize@url \@href}%
\providecommand \@href[1]{\@@startlink{#1}\@@href}%
\providecommand \@@href[1]{\endgroup#1\@@endlink}%
\providecommand \@sanitize@url [0]{\catcode `\\12\catcode `\$12\catcode
  `\&12\catcode `\#12\catcode `\^12\catcode `\_12\catcode `\%12\relax}%
\providecommand \@@startlink[1]{}%
\providecommand \@@endlink[0]{}%
\providecommand \url  [0]{\begingroup\@sanitize@url \@url }%
\providecommand \@url [1]{\endgroup\@href {#1}{\urlprefix }}%
\providecommand \urlprefix  [0]{URL }%
\providecommand \Eprint [0]{\href }%
\providecommand \doibase [0]{http://dx.doi.org/}%
\providecommand \selectlanguage [0]{\@gobble}%
\providecommand \bibinfo  [0]{\@secondoftwo}%
\providecommand \bibfield  [0]{\@secondoftwo}%
\providecommand \translation [1]{[#1]}%
\providecommand \BibitemOpen [0]{}%
\providecommand \bibitemStop [0]{}%
\providecommand \bibitemNoStop [0]{.\EOS\space}%
\providecommand \EOS [0]{\spacefactor3000\relax}%
\providecommand \BibitemShut  [1]{\csname bibitem#1\endcsname}%
\let\auto@bib@innerbib\@empty
%</preamble>
\bibitem [{\citenamefont {Arciprete}\ \emph {et~al.}(2013)\citenamefont
  {Arciprete}, \citenamefont {Placidi}, \citenamefont {Magri}, \citenamefont
  {Fanfoni}, \citenamefont {Balzarotti},\ and\ \citenamefont
  {Patella}}]{ArcipreteACSNano2013}%
  \BibitemOpen
  \bibfield  {author} {\bibinfo {author} {\bibfnamefont {F.}~\bibnamefont
  {Arciprete}}, \bibinfo {author} {\bibfnamefont {E.}~\bibnamefont {Placidi}},
  \bibinfo {author} {\bibfnamefont {R.}~\bibnamefont {Magri}}, \bibinfo
  {author} {\bibfnamefont {M.}~\bibnamefont {Fanfoni}}, \bibinfo {author}
  {\bibfnamefont {A.}~\bibnamefont {Balzarotti}}, \ and\ \bibinfo {author}
  {\bibfnamefont {F.}~\bibnamefont {Patella}},\ }\href@noop {} {\bibfield
  {journal} {\bibinfo  {journal} {ACS Nano}\ }\textbf {\bibinfo {volume} {7}},\
  \bibinfo {pages} {3868} (\bibinfo {year} {2013})}\BibitemShut {NoStop}%
\bibitem [{\citenamefont {Dvorak}\ \emph {et~al.}(2016)\citenamefont {Dvorak},
  \citenamefont {Farnesi~Camellone}, \citenamefont {Tovt}, \citenamefont
  {Tran}, \citenamefont {Negreiros}, \citenamefont {Vorokhta}, \citenamefont
  {Skala}, \citenamefont {Matolinova}, \citenamefont {Myslive{\"A}ek},
  \citenamefont {Matolin},\ and\ \citenamefont {Fabris}}]{Dvorak2016}%
  \BibitemOpen
  \bibfield  {author} {\bibinfo {author} {\bibfnamefont {F.}~\bibnamefont
  {Dvorak}}, \bibinfo {author} {\bibfnamefont {M.}~\bibnamefont
  {Farnesi~Camellone}}, \bibinfo {author} {\bibfnamefont {A.}~\bibnamefont
  {Tovt}}, \bibinfo {author} {\bibfnamefont {N.-D.}\ \bibnamefont {Tran}},
  \bibinfo {author} {\bibfnamefont {F.~R.}\ \bibnamefont {Negreiros}}, \bibinfo
  {author} {\bibfnamefont {M.}~\bibnamefont {Vorokhta}}, \bibinfo {author}
  {\bibfnamefont {T.}~\bibnamefont {Skala}}, \bibinfo {author} {\bibfnamefont
  {I.}~\bibnamefont {Matolinova}}, \bibinfo {author} {\bibfnamefont
  {J.}~\bibnamefont {Myslive{\"A}ek}}, \bibinfo {author} {\bibfnamefont
  {V.}~\bibnamefont {Matolin}}, \ and\ \bibinfo {author} {\bibfnamefont
  {S.}~\bibnamefont {Fabris}},\ }\href@noop {} {\ \textbf {\bibinfo {volume}
  {7}},\ \bibinfo {pages} {10801 EP } (\bibinfo {year} {2016})},\ \bibinfo
  {note} {article}\BibitemShut {NoStop}%
\bibitem [{\citenamefont {Miccio}\ \emph {et~al.}(2016)\citenamefont {Miccio},
  \citenamefont {Setvin}, \citenamefont {Müller}, \citenamefont {Abadía},
  \citenamefont {Piquero}, \citenamefont {Lobo-Checa}, \citenamefont
  {Schiller}, \citenamefont {Rogero}, \citenamefont {Schmid}, \citenamefont
  {Sánchez-Portal}, \citenamefont {Diebold},\ and\ \citenamefont
  {Ortega}}]{Miccio2016}%
  \BibitemOpen
  \bibfield  {author} {\bibinfo {author} {\bibfnamefont {L.~A.}\ \bibnamefont
  {Miccio}}, \bibinfo {author} {\bibfnamefont {M.}~\bibnamefont {Setvin}},
  \bibinfo {author} {\bibfnamefont {M.}~\bibnamefont {Müller}}, \bibinfo
  {author} {\bibfnamefont {M.}~\bibnamefont {Abadía}}, \bibinfo {author}
  {\bibfnamefont {I.}~\bibnamefont {Piquero}}, \bibinfo {author} {\bibfnamefont
  {J.}~\bibnamefont {Lobo-Checa}}, \bibinfo {author} {\bibfnamefont
  {F.}~\bibnamefont {Schiller}}, \bibinfo {author} {\bibfnamefont
  {C.}~\bibnamefont {Rogero}}, \bibinfo {author} {\bibfnamefont
  {M.}~\bibnamefont {Schmid}}, \bibinfo {author} {\bibfnamefont
  {D.}~\bibnamefont {Sánchez-Portal}}, \bibinfo {author} {\bibfnamefont
  {U.}~\bibnamefont {Diebold}}, \ and\ \bibinfo {author} {\bibfnamefont
  {J.~E.}\ \bibnamefont {Ortega}},\ }\href@noop {} {\bibfield  {journal}
  {\bibinfo  {journal} {Nano Letters}\ }\textbf {\bibinfo {volume} {16}},\
  \bibinfo {pages} {2017} (\bibinfo {year} {2016})},\ \bibinfo {note} {pMID:
  26752001}\BibitemShut {NoStop}%
\bibitem [{\citenamefont {Gruber}\ and\ \citenamefont
  {Mullins}(1967)}]{GruberJPhysChemSolids1967}%
  \BibitemOpen
  \bibfield  {author} {\bibinfo {author} {\bibfnamefont {E.}~\bibnamefont
  {Gruber}}\ and\ \bibinfo {author} {\bibfnamefont {W.}~\bibnamefont
  {Mullins}},\ }\href@noop {} {\bibfield  {journal} {\bibinfo  {journal}
  {Journal of Physics and Chemistry of Solids}\ }\textbf {\bibinfo {volume}
  {28}},\ \bibinfo {pages} {875 } (\bibinfo {year} {1967})}\BibitemShut
  {NoStop}%
\bibitem [{\citenamefont {Muller}\ and\ \citenamefont
  {Sa\'{u}l}(2004)}]{MullerSurfSciRep2004}%
  \BibitemOpen
  \bibfield  {author} {\bibinfo {author} {\bibfnamefont {P.}~\bibnamefont
  {Muller}}\ and\ \bibinfo {author} {\bibfnamefont {A.}~\bibnamefont
  {Sa\'{u}l}},\ }\href@noop {} {\bibfield  {journal} {\bibinfo  {journal}
  {Surface Science Reports}\ }\textbf {\bibinfo {volume} {54}},\ \bibinfo
  {pages} {157 } (\bibinfo {year} {2004})}\BibitemShut {NoStop}%
\bibitem [{\citenamefont {Frohn}\ \emph {et~al.}(1991)\citenamefont {Frohn},
  \citenamefont {Giesen}, \citenamefont {Poensgen}, \citenamefont {Wolf},\ and\
  \citenamefont {Ibach}}]{Frohn1991}%
  \BibitemOpen
  \bibfield  {author} {\bibinfo {author} {\bibfnamefont {J.}~\bibnamefont
  {Frohn}}, \bibinfo {author} {\bibfnamefont {M.}~\bibnamefont {Giesen}},
  \bibinfo {author} {\bibfnamefont {M.}~\bibnamefont {Poensgen}}, \bibinfo
  {author} {\bibfnamefont {J.~F.}\ \bibnamefont {Wolf}}, \ and\ \bibinfo
  {author} {\bibfnamefont {H.}~\bibnamefont {Ibach}},\ }\href@noop {}
  {\bibfield  {journal} {\bibinfo  {journal} {Phys. Rev. Lett.}\ }\textbf
  {\bibinfo {volume} {67}},\ \bibinfo {pages} {3543} (\bibinfo {year}
  {1991})}\BibitemShut {NoStop}%
\bibitem [{\citenamefont {Pai}\ \emph {et~al.}(1994)\citenamefont {Pai},
  \citenamefont {Ozcomert}, \citenamefont {Bartelt}, \citenamefont {Einstein},\
  and\ \citenamefont {Reutt-Robey}}]{PaiSurfSci1994}%
  \BibitemOpen
  \bibfield  {author} {\bibinfo {author} {\bibfnamefont {W.}~\bibnamefont
  {Pai}}, \bibinfo {author} {\bibfnamefont {J.}~\bibnamefont {Ozcomert}},
  \bibinfo {author} {\bibfnamefont {N.}~\bibnamefont {Bartelt}}, \bibinfo
  {author} {\bibfnamefont {T.}~\bibnamefont {Einstein}}, \ and\ \bibinfo
  {author} {\bibfnamefont {J.}~\bibnamefont {Reutt-Robey}},\ }\href@noop {}
  {\bibfield  {journal} {\bibinfo  {journal} {Surface Science}\ }\textbf
  {\bibinfo {volume} {307–309, Part B}},\ \bibinfo {pages} {747 } (\bibinfo
  {year} {1994})},\ \bibinfo {note} {proceedings of the European Conference on
  Surface Science}\BibitemShut {NoStop}%
\bibitem [{\citenamefont {Song}\ and\ \citenamefont
  {Mochrie}(1995)}]{SongPRB1995}%
  \BibitemOpen
  \bibfield  {author} {\bibinfo {author} {\bibfnamefont {S.}~\bibnamefont
  {Song}}\ and\ \bibinfo {author} {\bibfnamefont {S.~G.~J.}\ \bibnamefont
  {Mochrie}},\ }\href@noop {} {\bibfield  {journal} {\bibinfo  {journal} {Phys.
  Rev. B}\ }\textbf {\bibinfo {volume} {51}},\ \bibinfo {pages} {10068}
  (\bibinfo {year} {1995})}\BibitemShut {NoStop}%
\bibitem [{\citenamefont {Ciobanu}\ \emph {et~al.}(2003)\citenamefont
  {Ciobanu}, \citenamefont {Tambe}, \citenamefont {Shenoy}, \citenamefont
  {Wang},\ and\ \citenamefont {Ho}}]{CiobanuPRB2003}%
  \BibitemOpen
  \bibfield  {author} {\bibinfo {author} {\bibfnamefont {C.~V.}\ \bibnamefont
  {Ciobanu}}, \bibinfo {author} {\bibfnamefont {D.~T.}\ \bibnamefont {Tambe}},
  \bibinfo {author} {\bibfnamefont {V.~B.}\ \bibnamefont {Shenoy}}, \bibinfo
  {author} {\bibfnamefont {C.-Z.}\ \bibnamefont {Wang}}, \ and\ \bibinfo
  {author} {\bibfnamefont {K.-M.}\ \bibnamefont {Ho}},\ }\href@noop {}
  {\bibfield  {journal} {\bibinfo  {journal} {Phys. Rev. B}\ }\textbf {\bibinfo
  {volume} {68}},\ \bibinfo {pages} {201302} (\bibinfo {year}
  {2003})}\BibitemShut {NoStop}%
\bibitem [{\citenamefont {Redfield}\ and\ \citenamefont
  {Zangwill}(1992)}]{RedfieldPRB1992}%
  \BibitemOpen
  \bibfield  {author} {\bibinfo {author} {\bibfnamefont {A.~C.}\ \bibnamefont
  {Redfield}}\ and\ \bibinfo {author} {\bibfnamefont {A.}~\bibnamefont
  {Zangwill}},\ }\href@noop {} {\bibfield  {journal} {\bibinfo  {journal}
  {Phys. Rev. B}\ }\textbf {\bibinfo {volume} {46}},\ \bibinfo {pages} {4289}
  (\bibinfo {year} {1992})}\BibitemShut {NoStop}%
\bibitem [{\citenamefont {Lelarge}\ \emph {et~al.}(1997)\citenamefont
  {Lelarge}, \citenamefont {Wang}, \citenamefont {Cavanna}, \citenamefont
  {Laruelle},\ and\ \citenamefont {Etienne}}]{LelargeEPL1997}%
  \BibitemOpen
  \bibfield  {author} {\bibinfo {author} {\bibfnamefont {F.}~\bibnamefont
  {Lelarge}}, \bibinfo {author} {\bibfnamefont {Z.~Z.}\ \bibnamefont {Wang}},
  \bibinfo {author} {\bibfnamefont {A.}~\bibnamefont {Cavanna}}, \bibinfo
  {author} {\bibfnamefont {F.}~\bibnamefont {Laruelle}}, \ and\ \bibinfo
  {author} {\bibfnamefont {B.}~\bibnamefont {Etienne}},\ }\href
  {http://stacks.iop.org/0295-5075/39/i=1/a=097} {\bibfield  {journal}
  {\bibinfo  {journal} {EPL (Europhysics Letters)}\ }\textbf {\bibinfo {volume}
  {39}},\ \bibinfo {pages} {97} (\bibinfo {year} {1997})}\BibitemShut {NoStop}%
\bibitem [{\citenamefont {Magri}\ \emph {et~al.}(2014)\citenamefont {Magri},
  \citenamefont {Gupta},\ and\ \citenamefont {Rosini}}]{Magri2014}%
  \BibitemOpen
  \bibfield  {author} {\bibinfo {author} {\bibfnamefont {R.}~\bibnamefont
  {Magri}}, \bibinfo {author} {\bibfnamefont {S.~K.}\ \bibnamefont {Gupta}}, \
  and\ \bibinfo {author} {\bibfnamefont {M.}~\bibnamefont {Rosini}},\
  }\href@noop {} {\bibfield  {journal} {\bibinfo  {journal} {Phys. Rev. B}\
  }\textbf {\bibinfo {volume} {90}},\ \bibinfo {pages} {115314} (\bibinfo
  {year} {2014})}\BibitemShut {NoStop}%
\bibitem [{\citenamefont {Magri}\ \emph {et~al.}(2016)\citenamefont {Magri},
  \citenamefont {Gupta},\ and\ \citenamefont {Rosini}}]{Erratum}%
  \BibitemOpen
  \bibfield  {author} {\bibinfo {author} {\bibfnamefont {R.}~\bibnamefont
  {Magri}}, \bibinfo {author} {\bibfnamefont {S.~K.}\ \bibnamefont {Gupta}}, \
  and\ \bibinfo {author} {\bibfnamefont {M.}~\bibnamefont {Rosini}},\
  }\href@noop {} {\bibfield  {journal} {\bibinfo  {journal} {Phys. Rev. B}\
  }\textbf {\bibinfo {volume} {94}},\ \bibinfo {pages} {239909} (\bibinfo
  {year} {2016})}\BibitemShut {NoStop}%
\bibitem [{\citenamefont {Heller}\ \emph {et~al.}(1993)\citenamefont {Heller},
  \citenamefont {Zhang},\ and\ \citenamefont {Lagally}}]{HellerPRL1993}%
  \BibitemOpen
  \bibfield  {author} {\bibinfo {author} {\bibfnamefont {E.~J.}\ \bibnamefont
  {Heller}}, \bibinfo {author} {\bibfnamefont {Z.~Y.}\ \bibnamefont {Zhang}}, \
  and\ \bibinfo {author} {\bibfnamefont {M.~G.}\ \bibnamefont {Lagally}},\
  }\href@noop {} {\bibfield  {journal} {\bibinfo  {journal} {Phys. Rev. Lett.}\
  }\textbf {\bibinfo {volume} {71}},\ \bibinfo {pages} {743} (\bibinfo {year}
  {1993})}\BibitemShut {NoStop}%
\bibitem [{\citenamefont {Kanisawa}\ \emph {et~al.}(1996)\citenamefont
  {Kanisawa}, \citenamefont {Yamaguchi},\ and\ \citenamefont
  {Horikoshi}}]{KanisawaPRB1996}%
  \BibitemOpen
  \bibfield  {author} {\bibinfo {author} {\bibfnamefont {K.}~\bibnamefont
  {Kanisawa}}, \bibinfo {author} {\bibfnamefont {H.}~\bibnamefont {Yamaguchi}},
  \ and\ \bibinfo {author} {\bibfnamefont {Y.}~\bibnamefont {Horikoshi}},\
  }\href@noop {} {\bibfield  {journal} {\bibinfo  {journal} {Phys. Rev. B}\
  }\textbf {\bibinfo {volume} {54}},\ \bibinfo {pages} {4428} (\bibinfo {year}
  {1996})}\BibitemShut {NoStop}%
\bibitem [{\citenamefont {Pashley}\ \emph {et~al.}(1991)\citenamefont
  {Pashley}, \citenamefont {Haberern},\ and\ \citenamefont
  {Gaines}}]{PashleyAPL1991}%
  \BibitemOpen
  \bibfield  {author} {\bibinfo {author} {\bibfnamefont {M.~D.}\ \bibnamefont
  {Pashley}}, \bibinfo {author} {\bibfnamefont {K.~W.}\ \bibnamefont
  {Haberern}}, \ and\ \bibinfo {author} {\bibfnamefont {J.~M.}\ \bibnamefont
  {Gaines}},\ }\href@noop {} {\bibfield  {journal} {\bibinfo  {journal}
  {Applied Physics Letters}\ }\textbf {\bibinfo {volume} {58}},\ \bibinfo
  {pages} {406} (\bibinfo {year} {1991})}\BibitemShut {NoStop}%
\bibitem [{\citenamefont {Giannozzi}\ \emph {et~al.}(2009)\citenamefont
  {Giannozzi}, \citenamefont {Baroni}, \citenamefont {Bonini}, \citenamefont
  {Calandra}, \citenamefont {Car}, \citenamefont {Cavazzoni}, \citenamefont
  {Ceresoli}, \citenamefont {Chiarotti}, \citenamefont {Cococcioni},
  \citenamefont {Dabo}, \citenamefont {{Dal Corso}}, \citenamefont
  {de~Gironcoli}, \citenamefont {Fabris}, \citenamefont {Fratesi},
  \citenamefont {Gebauer}, \citenamefont {Gerstmann}, \citenamefont
  {Gougoussis}, \citenamefont {Kokalj}, \citenamefont {Lazzeri}, \citenamefont
  {Martin-Samos}, \citenamefont {Marzari}, \citenamefont {Mauri}, \citenamefont
  {Mazzarello}, \citenamefont {Paolini}, \citenamefont {Pasquarello},
  \citenamefont {Paulatto}, \citenamefont {Sbraccia}, \citenamefont {Scandolo},
  \citenamefont {Sclauzero}, \citenamefont {Seitsonen}, \citenamefont
  {Smogunov}, \citenamefont {Umari},\ and\ \citenamefont
  {Wentzcovitch}}]{QE-2009}%
  \BibitemOpen
  \bibfield  {author} {\bibinfo {author} {\bibfnamefont {P.}~\bibnamefont
  {Giannozzi}}, \bibinfo {author} {\bibfnamefont {S.}~\bibnamefont {Baroni}},
  \bibinfo {author} {\bibfnamefont {N.}~\bibnamefont {Bonini}}, \bibinfo
  {author} {\bibfnamefont {M.}~\bibnamefont {Calandra}}, \bibinfo {author}
  {\bibfnamefont {R.}~\bibnamefont {Car}}, \bibinfo {author} {\bibfnamefont
  {C.}~\bibnamefont {Cavazzoni}}, \bibinfo {author} {\bibfnamefont
  {D.}~\bibnamefont {Ceresoli}}, \bibinfo {author} {\bibfnamefont {G.~L.}\
  \bibnamefont {Chiarotti}}, \bibinfo {author} {\bibfnamefont {M.}~\bibnamefont
  {Cococcioni}}, \bibinfo {author} {\bibfnamefont {I.}~\bibnamefont {Dabo}},
  \bibinfo {author} {\bibfnamefont {A.}~\bibnamefont {{Dal Corso}}}, \bibinfo
  {author} {\bibfnamefont {S.}~\bibnamefont {de~Gironcoli}}, \bibinfo {author}
  {\bibfnamefont {S.}~\bibnamefont {Fabris}}, \bibinfo {author} {\bibfnamefont
  {G.}~\bibnamefont {Fratesi}}, \bibinfo {author} {\bibfnamefont
  {R.}~\bibnamefont {Gebauer}}, \bibinfo {author} {\bibfnamefont
  {U.}~\bibnamefont {Gerstmann}}, \bibinfo {author} {\bibfnamefont
  {C.}~\bibnamefont {Gougoussis}}, \bibinfo {author} {\bibfnamefont
  {A.}~\bibnamefont {Kokalj}}, \bibinfo {author} {\bibfnamefont
  {M.}~\bibnamefont {Lazzeri}}, \bibinfo {author} {\bibfnamefont
  {L.}~\bibnamefont {Martin-Samos}}, \bibinfo {author} {\bibfnamefont
  {N.}~\bibnamefont {Marzari}}, \bibinfo {author} {\bibfnamefont
  {F.}~\bibnamefont {Mauri}}, \bibinfo {author} {\bibfnamefont
  {R.}~\bibnamefont {Mazzarello}}, \bibinfo {author} {\bibfnamefont
  {S.}~\bibnamefont {Paolini}}, \bibinfo {author} {\bibfnamefont
  {A.}~\bibnamefont {Pasquarello}}, \bibinfo {author} {\bibfnamefont
  {L.}~\bibnamefont {Paulatto}}, \bibinfo {author} {\bibfnamefont
  {C.}~\bibnamefont {Sbraccia}}, \bibinfo {author} {\bibfnamefont
  {S.}~\bibnamefont {Scandolo}}, \bibinfo {author} {\bibfnamefont
  {G.}~\bibnamefont {Sclauzero}}, \bibinfo {author} {\bibfnamefont {A.~P.}\
  \bibnamefont {Seitsonen}}, \bibinfo {author} {\bibfnamefont {A.}~\bibnamefont
  {Smogunov}}, \bibinfo {author} {\bibfnamefont {P.}~\bibnamefont {Umari}}, \
  and\ \bibinfo {author} {\bibfnamefont {R.~M.}\ \bibnamefont {Wentzcovitch}},\
  }\href@noop {} {\bibfield  {journal} {\bibinfo  {journal} {Journal of
  Physics: Condensed Matter}\ }\textbf {\bibinfo {volume} {21}},\ \bibinfo
  {pages} {395502 (19pp)} (\bibinfo {year} {2009})}\BibitemShut {NoStop}%
\bibitem [{\citenamefont {Giannozzi}\ \emph {et~al.}(2017)\citenamefont
  {Giannozzi}, \citenamefont {Andreussi}, \citenamefont {Brumme}, \citenamefont
  {Bunau}, \citenamefont {Nardelli}, \citenamefont {Calandra}, \citenamefont
  {Car}, \citenamefont {Cavazzoni}, \citenamefont {Ceresoli}, \citenamefont
  {Cococcioni}, \citenamefont {Colonna}, \citenamefont {Carnimeo},
  \citenamefont {Corso}, \citenamefont {de~Gironcoli}, \citenamefont {Delugas},
  \citenamefont {Jr}, \citenamefont {Ferretti}, \citenamefont {Floris},
  \citenamefont {Fratesi}, \citenamefont {Fugallo}, \citenamefont {Gebauer},
  \citenamefont {Gerstmann}, \citenamefont {Giustino}, \citenamefont {Gorni},
  \citenamefont {Jia}, \citenamefont {Kawamura}, \citenamefont {Ko},
  \citenamefont {Kokalj}, \citenamefont {Küçükbenli}, \citenamefont {Lazzeri},
  \citenamefont {Marsili}, \citenamefont {Marzari}, \citenamefont {Mauri},
  \citenamefont {Nguyen}, \citenamefont {Nguyen}, \citenamefont {de-la Roza},
  \citenamefont {Paulatto}, \citenamefont {Poncé}, \citenamefont {Rocca},
  \citenamefont {Sabatini}, \citenamefont {Santra}, \citenamefont {Schlipf},
  \citenamefont {Seitsonen}, \citenamefont {Smogunov}, \citenamefont {Timrov},
  \citenamefont {Thonhauser}, \citenamefont {Umari}, \citenamefont {Vast},
  \citenamefont {Wu},\ and\ \citenamefont {Baroni}}]{QE-2017}%
  \BibitemOpen
  \bibfield  {author} {\bibinfo {author} {\bibfnamefont {P.}~\bibnamefont
  {Giannozzi}}, \bibinfo {author} {\bibfnamefont {O.}~\bibnamefont
  {Andreussi}}, \bibinfo {author} {\bibfnamefont {T.}~\bibnamefont {Brumme}},
  \bibinfo {author} {\bibfnamefont {O.}~\bibnamefont {Bunau}}, \bibinfo
  {author} {\bibfnamefont {M.~B.}\ \bibnamefont {Nardelli}}, \bibinfo {author}
  {\bibfnamefont {M.}~\bibnamefont {Calandra}}, \bibinfo {author}
  {\bibfnamefont {R.}~\bibnamefont {Car}}, \bibinfo {author} {\bibfnamefont
  {C.}~\bibnamefont {Cavazzoni}}, \bibinfo {author} {\bibfnamefont
  {D.}~\bibnamefont {Ceresoli}}, \bibinfo {author} {\bibfnamefont
  {M.}~\bibnamefont {Cococcioni}}, \bibinfo {author} {\bibfnamefont
  {N.}~\bibnamefont {Colonna}}, \bibinfo {author} {\bibfnamefont
  {I.}~\bibnamefont {Carnimeo}}, \bibinfo {author} {\bibfnamefont {A.~D.}\
  \bibnamefont {Corso}}, \bibinfo {author} {\bibfnamefont {S.}~\bibnamefont
  {de~Gironcoli}}, \bibinfo {author} {\bibfnamefont {P.}~\bibnamefont
  {Delugas}}, \bibinfo {author} {\bibfnamefont {R.~A.~D.}\ \bibnamefont {Jr}},
  \bibinfo {author} {\bibfnamefont {A.}~\bibnamefont {Ferretti}}, \bibinfo
  {author} {\bibfnamefont {A.}~\bibnamefont {Floris}}, \bibinfo {author}
  {\bibfnamefont {G.}~\bibnamefont {Fratesi}}, \bibinfo {author} {\bibfnamefont
  {G.}~\bibnamefont {Fugallo}}, \bibinfo {author} {\bibfnamefont
  {R.}~\bibnamefont {Gebauer}}, \bibinfo {author} {\bibfnamefont
  {U.}~\bibnamefont {Gerstmann}}, \bibinfo {author} {\bibfnamefont
  {F.}~\bibnamefont {Giustino}}, \bibinfo {author} {\bibfnamefont
  {T.}~\bibnamefont {Gorni}}, \bibinfo {author} {\bibfnamefont
  {J.}~\bibnamefont {Jia}}, \bibinfo {author} {\bibfnamefont {M.}~\bibnamefont
  {Kawamura}}, \bibinfo {author} {\bibfnamefont {H.-Y.}\ \bibnamefont {Ko}},
  \bibinfo {author} {\bibfnamefont {A.}~\bibnamefont {Kokalj}}, \bibinfo
  {author} {\bibfnamefont {E.}~\bibnamefont {Küçükbenli}}, \bibinfo {author}
  {\bibfnamefont {M.}~\bibnamefont {Lazzeri}}, \bibinfo {author} {\bibfnamefont
  {M.}~\bibnamefont {Marsili}}, \bibinfo {author} {\bibfnamefont
  {N.}~\bibnamefont {Marzari}}, \bibinfo {author} {\bibfnamefont
  {F.}~\bibnamefont {Mauri}}, \bibinfo {author} {\bibfnamefont {N.~L.}\
  \bibnamefont {Nguyen}}, \bibinfo {author} {\bibfnamefont {H.-V.}\
  \bibnamefont {Nguyen}}, \bibinfo {author} {\bibfnamefont {A.~O.}\
  \bibnamefont {de-la Roza}}, \bibinfo {author} {\bibfnamefont
  {L.}~\bibnamefont {Paulatto}}, \bibinfo {author} {\bibfnamefont
  {S.}~\bibnamefont {Poncé}}, \bibinfo {author} {\bibfnamefont
  {D.}~\bibnamefont {Rocca}}, \bibinfo {author} {\bibfnamefont
  {R.}~\bibnamefont {Sabatini}}, \bibinfo {author} {\bibfnamefont
  {B.}~\bibnamefont {Santra}}, \bibinfo {author} {\bibfnamefont
  {M.}~\bibnamefont {Schlipf}}, \bibinfo {author} {\bibfnamefont {A.~P.}\
  \bibnamefont {Seitsonen}}, \bibinfo {author} {\bibfnamefont {A.}~\bibnamefont
  {Smogunov}}, \bibinfo {author} {\bibfnamefont {I.}~\bibnamefont {Timrov}},
  \bibinfo {author} {\bibfnamefont {T.}~\bibnamefont {Thonhauser}}, \bibinfo
  {author} {\bibfnamefont {P.}~\bibnamefont {Umari}}, \bibinfo {author}
  {\bibfnamefont {N.}~\bibnamefont {Vast}}, \bibinfo {author} {\bibfnamefont
  {X.}~\bibnamefont {Wu}}, \ and\ \bibinfo {author} {\bibfnamefont
  {S.}~\bibnamefont {Baroni}},\ }\href
  {http://stacks.iop.org/0953-8984/29/i=46/a=465901} {\bibfield  {journal}
  {\bibinfo  {journal} {Journal of Physics: Condensed Matter}\ }\textbf
  {\bibinfo {volume} {29}},\ \bibinfo {pages} {465901} (\bibinfo {year}
  {2017})}\BibitemShut {NoStop}%
\bibitem [{\citenamefont {Hardcastle}\ \emph {et~al.}(2013)\citenamefont
  {Hardcastle}, \citenamefont {Seabourne}, \citenamefont {Brydson},
  \citenamefont {Livi},\ and\ \citenamefont {Scott}}]{Hardcastle2013}%
  \BibitemOpen
  \bibfield  {author} {\bibinfo {author} {\bibfnamefont {T.~P.}\ \bibnamefont
  {Hardcastle}}, \bibinfo {author} {\bibfnamefont {C.~R.}\ \bibnamefont
  {Seabourne}}, \bibinfo {author} {\bibfnamefont {R.~M.~D.}\ \bibnamefont
  {Brydson}}, \bibinfo {author} {\bibfnamefont {K.~J.~T.}\ \bibnamefont
  {Livi}}, \ and\ \bibinfo {author} {\bibfnamefont {A.~J.}\ \bibnamefont
  {Scott}},\ }\href {\doibase 10.1021/jp4078135} {\bibfield  {journal}
  {\bibinfo  {journal} {The Journal of Physical Chemistry C}\ }\textbf
  {\bibinfo {volume} {117}},\ \bibinfo {pages} {23766} (\bibinfo {year}
  {2013})},\ \Eprint {http://arxiv.org/abs/https://doi.org/10.1021/jp4078135}
  {https://doi.org/10.1021/jp4078135} \BibitemShut {NoStop}%
\bibitem [{\citenamefont {Zhang}\ \emph {et~al.}(2015)\citenamefont {Zhang},
  \citenamefont {Cui}, \citenamefont {Griffiths}, \citenamefont {Fu},
  \citenamefont {Freysoldt}, \citenamefont {Neugebauer}, \citenamefont
  {Humphreys},\ and\ \citenamefont {Oliver}}]{Zhang2015}%
  \BibitemOpen
  \bibfield  {author} {\bibinfo {author} {\bibfnamefont {S.}~\bibnamefont
  {Zhang}}, \bibinfo {author} {\bibfnamefont {Y.}~\bibnamefont {Cui}}, \bibinfo
  {author} {\bibfnamefont {J.~T.}\ \bibnamefont {Griffiths}}, \bibinfo {author}
  {\bibfnamefont {W.~Y.}\ \bibnamefont {Fu}}, \bibinfo {author} {\bibfnamefont
  {C.}~\bibnamefont {Freysoldt}}, \bibinfo {author} {\bibfnamefont
  {J.}~\bibnamefont {Neugebauer}}, \bibinfo {author} {\bibfnamefont {C.~J.}\
  \bibnamefont {Humphreys}}, \ and\ \bibinfo {author} {\bibfnamefont {R.~A.}\
  \bibnamefont {Oliver}},\ }\href {\doibase 10.1103/PhysRevB.92.245202}
  {\bibfield  {journal} {\bibinfo  {journal} {Phys. Rev. B}\ }\textbf {\bibinfo
  {volume} {92}},\ \bibinfo {pages} {245202} (\bibinfo {year}
  {2015})}\BibitemShut {NoStop}%
\bibitem [{\citenamefont {King-Smith}\ and\ \citenamefont
  {Vanderbilt}(1994)}]{King-Smith1994}%
  \BibitemOpen
  \bibfield  {author} {\bibinfo {author} {\bibfnamefont {R.~D.}\ \bibnamefont
  {King-Smith}}\ and\ \bibinfo {author} {\bibfnamefont {D.}~\bibnamefont
  {Vanderbilt}},\ }\href {\doibase 10.1103/PhysRevB.49.5828} {\bibfield
  {journal} {\bibinfo  {journal} {Phys. Rev. B}\ }\textbf {\bibinfo {volume}
  {49}},\ \bibinfo {pages} {5828} (\bibinfo {year} {1994})}\BibitemShut
  {NoStop}%
\bibitem [{\citenamefont {Kresse}\ \emph {et~al.}(1995)\citenamefont {Kresse},
  \citenamefont {Furthmüller},\ and\ \citenamefont {Hafner}}]{Kresse1995}%
  \BibitemOpen
  \bibfield  {author} {\bibinfo {author} {\bibfnamefont {G.}~\bibnamefont
  {Kresse}}, \bibinfo {author} {\bibfnamefont {J.}~\bibnamefont
  {Furthmüller}}, \ and\ \bibinfo {author} {\bibfnamefont {J.}~\bibnamefont
  {Hafner}},\ }\href {http://stacks.iop.org/0295-5075/32/i=9/a=005} {\bibfield
  {journal} {\bibinfo  {journal} {EPL (Europhysics Letters)}\ }\textbf
  {\bibinfo {volume} {32}},\ \bibinfo {pages} {729} (\bibinfo {year}
  {1995})}\BibitemShut {NoStop}%
\bibitem [{\citenamefont {Fletcher}(1987)}]{fletcher1987}%
  \BibitemOpen
  \bibfield  {author} {\bibinfo {author} {\bibfnamefont {R.}~\bibnamefont
  {Fletcher}},\ }\href@noop {} {\bibfield  {journal} {\bibinfo  {journal} {John
  Wiley \& Sons}\ }\textbf {\bibinfo {volume} {80}} (\bibinfo {year}
  {1987})}\BibitemShut {NoStop}%
\bibitem [{\citenamefont {Billeter}\ \emph {et~al.}(2003)\citenamefont
  {Billeter}, \citenamefont {Curioni},\ and\ \citenamefont
  {Andreoni}}]{billeter2003}%
  \BibitemOpen
  \bibfield  {author} {\bibinfo {author} {\bibfnamefont {S.~R.}\ \bibnamefont
  {Billeter}}, \bibinfo {author} {\bibfnamefont {A.}~\bibnamefont {Curioni}}, \
  and\ \bibinfo {author} {\bibfnamefont {W.}~\bibnamefont {Andreoni}},\
  }\href@noop {} {\bibfield  {journal} {\bibinfo  {journal} {Computational
  materials science}\ }\textbf {\bibinfo {volume} {27}},\ \bibinfo {pages}
  {437} (\bibinfo {year} {2003})}\BibitemShut {NoStop}%
\bibitem [{\citenamefont {Marchenko}\ and\ \citenamefont
  {Parshin}(1980)}]{MarchenkoSovPhysJETP1980}%
  \BibitemOpen
  \bibfield  {author} {\bibinfo {author} {\bibfnamefont {V.~I.}\ \bibnamefont
  {Marchenko}}\ and\ \bibinfo {author} {\bibfnamefont {A.~Y.}\ \bibnamefont
  {Parshin}},\ }\href@noop {} {\bibfield  {journal} {\bibinfo  {journal} {Sov.
  Phys. JETP}\ }\textbf {\bibinfo {volume} {52}},\ \bibinfo {pages} {129}
  (\bibinfo {year} {1980})}\BibitemShut {NoStop}%
\end{thebibliography}%

%\begin{thebibliography}{9}

%\end{thebibliography}

%
\end{document}